\documentclass[aps,prd,amsmath,showpacs,nofootinbib]{revtex4}
\usepackage{graphicx}
\usepackage[cp1251]{inputenc}
\usepackage[english,russian]{babel}
\inputencoding{cp1251}

\newcommand{\beq}{\begin{equation}}
\newcommand{\eeq}{\end{equation}}

\newcommand{\bea}{\begin{eqnarray}}
\newcommand{\eea}{\end{eqnarray}}

\newcommand{\bs}{\begin{subequations}}
\newcommand{\es}{\end{subequations}}

\newcommand{\Ref}[1]{(\ref{#1})}
\newcommand{\pd}{\partial}

\begin{document}

\title{Slowly rotating scalar field wormholes: the second order
approximation}

\author{P.E. Kashargin$^a$}

\author{S.V. Sushkov$^{a,b}$}
\email{sergey_sushkov@mail.ru; sergey.sushkov@ksu.ru}
\affiliation{$^a$Department of General Relativity and Gravitation,
Kazan State University, Kremlevskaya str. 18, Kazan 420008,
Russia} %
\affiliation{$^b$Department of Mathematics, Tatar State
University of Humanities and Education, Tatarstan str. 2, Kazan
420021, Russia}

\date{\today}

\begin{abstract}
We discuss rotating wormholes in general relativity with a scalar
field with negative kinetic energy. To solve the problem, we use
the assumption about slow rotation. The role of a small
dimensionless parameter plays the ratio of the linear velocity of
rotation of the wormhole's throat and the velocity of light. We
construct the rotating wormhole solution in the second order
approximation with respect to the small parameter. The analysis
shows that the asymptotical mass of the rotating wormhole is
greater than that of the non-rotating one, and the NEC violation
in the rotating wormhole spacetime is weaker than that in the
non-rotating one.
\end{abstract}


\maketitle

\section{Introduction}
Wormholes are usually defined as topological handles in spacetime
linking widely separated regions of a single universe, or
``bridges'' joining two different spacetimes
\cite{MorTho,VisserBook}. As is well-known \cite{HocVis}, they can
exist {only if} their throats contain exotic matter which
possesses a negative pressure and violates the null energy
condition. The search of realistic physical models providing the
wormhole existence represents an important direction in wormhole
physics. Various models of this kind include scalar fields,
wormhole solutions in semiclassical gravity,
solutions in Brans-Dicke theory,
wormholes on branes,
wormholes supported by matter with exotic equations of state, such
as phantom energy, 
the Chaplygin gas, 
tachyon matter, 
and others \cite{footnote,LoboReview}.

It is worth being noticed that most of the investigations deal
with static spherically symmetric wormholes because of their
simplicity and high symmetry. At the same time it would be
important and interesting from the physical point of view to study
more wide classes of wormholes including non-static and rotating
ones. Non-static wormholes whose geometry is depending on time
have been discussed in the literature. In 1993 Roman \cite{Rom:93}
explored the possibility that inflation might provide a mechanism
for the enlargement of submicroscopic, i.e., Planck scale
wormholes to macroscopic size. He used the line element with the
exponential scale factor. Kim \cite{Kim:96} generalized the
Roman's consideration by using the scale factor in a general form.
Various aspects of non-static wormholes conformally related to
static wormhole geometries were investigated in
\cite{Anc-etal,Kar,KarSah,WanLet}; in particular, some issues
concerning WEC violation and traversability in these
time-dependent geometries were discussed. Kuhfittig \cite{Kuh:02}
considered a spherically symmetric wormhole spacetime with the
metric whose components are time-depending. Exact solutions
describing cosmological evolution of scalar field wormholes were
obtained in \cite{SusKim,SusZha}.

Rotating wormholes have also been an object for study. Some
general geometrical properties of stationary rotating wormholes
have been first analyzed by Teo \cite{Teo}. General requirements
for the stress-energy tensor necessary to generate the rotating
wormhole were discussed in \cite{BerHib}. The WEC violation and
traversability in the rotating wormhole spacetime were in details
studied in \cite{Kuh}. Kim \cite{Kim} investigated scalar
perturbations in a particular model of the rotating wormhole. In
Ref. \cite{JamRas} the authors studied a slowly rotating wormhole
surrounded by a cloud of charged particles. Arguments in favour of
the possibility of existence of semiclassical rotating wormholes
were given in \cite{Kha}.

In previous work \cite{KasSus} we were continuing the study of
rotating wormholes. Our aim was to construct an exact solution
describing these objects in general relativity with a scalar
field. As is well-known (see \cite{Ell,Bro}) a scalar ghost, i.e.
a scalar field with negative kinetic energy can support static
spherically symmetric wormholes. Moreover, such wormholes are
stable against linear spherically symmetric perturbations
\cite{stability}. In \cite{KasSus} we looked for rotating
wormholes supported by the scalar field with negative kinetic
energy. To solve the problem we supposed that a wormhole is very
slowly rotating and constructed a solution in the first order
approximation with respect to a small parameter characterizing the
velocity of rotation. In this approximation the only term
$\sim\Omega dtd\varphi$ is added to the initial non-rotating
wormhole metric, where $\Omega$ is the local angular velocity of
rotation. This term results in the well-know dragging effect in
general relativity. Namely, it was shown that a test particle
initially propagating along the radial direction turns out to be
involving into the wormhole rotation so that after passing through
the throat of a wormhole it continues its motion along a spiral
trajectory moving away from the throat. Propagation of light
exhibits similar behavior. The ray of light after passing through
the rotating wormhole throat is propagating along the spiral. At
the same time, the first order approximation does not give an
answer to a number of interesting and important problems
concerning the rotating wormhole mass, the NEC violation and
others. For this reason, in this paper we construct and analyze
the second order solution describing the rotating wormhole in
general relativity with the scalar field with negative kinetic
energy.

The paper is organized as follows. In Section \ref{genform} we
give some general formulas and write down the field equations. A
static spherically symmetric wormhole is briefly discussed in
Section \ref{statwh}. In Section \ref{rotwh} we formulate the
condition of slow rotation and introduce a small parameter
characterizing the rotation velocity. Then we construct a solution
describing a rotating wormhole in the second order approximation
with respect to this small parameter. Some properties of the
solution is analyzed in Section \ref{analysis}. Namely, we discuss
how a mass and a value of the NEC violation by a rotating wormhole
differ from those of a non-rotating one. A summary of results
obtained is given in Section \ref{conc}. In Appendix we give
details of solving field equations in the second order
approximation and write their solution in an explicit form.

\section{General formulas\label{genform}}
Consider general relativity with a scalar field $\Phi$, describing
by the action
\beq\label{action}
S=\int d^4x\sqrt{-g}\left[{\cal R} + (\nabla\Phi)^2\right],
\eeq
where $g_{\mu\nu}$ is a metric, $g=\det(g_{\mu\nu})$, $\cal R$ is
the scalar curvature, and
$(\nabla\Phi)^2=g^{\mu\nu}\Phi_{,\mu}\Phi_{,\nu}$ is the kinetic
term. Throughout the paper we use units $G=c=1$ and the signature
$(-+++)$. For this signature the $+$ sign  before the kinetic term
corresponds to negative kinetic energy, hence $\Phi$ is a {\em
ghost}.

Varying the action \Ref{action} with respect to $g_{\mu\nu}$ and
$\phi$ yields Einstein equations and the equation of motion of the
scalar field, respectively:
\bea\label{einstein}
&&{\cal R}_{\mu\nu}=-\Phi_{,\mu}\Phi_{,\nu},\\
\label{eqmo} &&\nabla^{\alpha}\nabla_{\alpha}\Phi=0.
\eea

In the paper we will search for solutions of the system
\Ref{einstein},\Ref{eqmo} describing rotating wormholes. A
spacetime with the stationary rotation possesses the axial
symmetry. As is known (see, e.g. \cite{Hartle}) a general axially
symmetric metric can be given in the following form:
\beq\label{metric}
ds^2=-Adt^2+Bdr^2+R^2[d\theta^2+\sin^2\theta(d\varphi-\Omega
dt)^2],
\eeq
where $A$, $B$, $R$, $\Omega$ are functions of $r$, $\theta$. The
function $\Omega$ has an explicit physical sense; it represents an
angular velocity of rotation in a point $(r,\theta)$. The
requirement of finiteness of the angular momentum $J$ measured by
a distant observer yields the following asymptotical condition for
$\Omega$ \cite{Land}:
\beq\label{as}
\Omega=\frac{2J}{r^3}+O(r^{-4})\quad {\rm as}\quad r\to\infty.
\eeq
Also, requiring that a spacetime should be asymptotically flat we
have $A\to1$, $B\to1$, and $R^2\to r^2$ as $r\to\infty$.

Eqs. \Ref{einstein}, \Ref{eqmo}, written for the metric
\Ref{metric}, are second-order partial differential equations for
five functions $A$, $B$, $R$, $\Omega$, and $\Phi$. Solving these
equations in a general form is a rather complicated mathematical
problem. To simplify the problem, in what follows we restrict
ourselves by the case of slow rotation.

\section{Static spherically symmetric wormhole\label{statwh}}
To formulate the condition of slow rotation, first of all we will
discuss the static spherically symmetric case. A static
spherically symmetric solution in general relativity with a ghost
scalar field was first found by Ellis \cite{Ell} and independently
by Bronnikov \cite{Bro}. This solution can be presented as follows
(see \cite{SusKim}):
\bea\label{stat.m}
&ds^2=-e^{2u(r)}dt^2+e^{-2u(r)}[dr^{2}+(r^{2}+a^2)(d\theta^2+\sin^2\theta
d\varphi^2 )],&\\
\label{stat.phi} &\displaystyle
\Phi(r)={\frac{(m^2+a^2)^{1/2}}{2\pi^{1/2}\, m}}\,u(r),&
\eea
where the radial coordinate $r$ varies from $-\infty$ to $\infty$,
$m$ and $a$ are free parameters, and
\beq\label{u}
u(r)=\frac{m}{a}\left(\arctan\frac{r}{a}-\frac{\pi}{2}\right).
\eeq
Taking into account the following asymptotical behavior:
\bea
e^{2u}|_{r\to\infty}&=&1-\frac{2m}{r}+O(r^{-2}), \nonumber\\
e^{2u}|_{r\to-\infty}&=&e^{-2\pi
m/a}\left(1-\frac{2m}{r}\right)+O(r^{-2}), \nonumber
\eea
we can see that the spacetime with the metric \Ref{stat.m}
possesses two asymptotically flat regions. The parameter $m$ plays
a role of the asymptotical mass for a distant observer located at
$r=\infty$. We will assume that $m\ge0$. The asymptotically flat
regions are connected by a throat whose radius corresponds to a
minimum of the radius of two-dimensional sphere,
$R^2(r)=e^{-2u(r)}(r^2+a^2)$. The minimum of $R(r)$ is achieved at
$r_{\rm th}=m$. The value $R_{\rm th}=R(r_{\rm th})$ is called the
radius of wormhole throat. It is worth noting that there exist
massless wormholes with $m=0$. In this case the metric
\Ref{stat.m} takes the especially simple form:
\beq\label{MTwh}
ds^2=-dt^2+dr^2+(r^2+a^2)(d\theta^2+\sin^2\theta d\varphi^2 ).
\eeq
It is interesting that the metric \Ref{MTwh} was proposed {a
priori} by Morris and Thorne in \cite{MorTho} as a simple example
of the wormhole spacetime metric.

\section{Rotating wormhole\label{rotwh}}
Now let us consider rotating wormholes. For this aim we take the
wormhole metric in the form \Ref{metric} with
$r\in(-\infty,+\infty)$. Assume that the throat of a wormhole
corresponds to the value $r=r_{\rm th}$. Define the throat's
radius as $R_{\rm th}=R|_{r=r_{\rm th},\theta=\pi/2}$. Also we
introduce the value $ \Omega_{\rm th}=\Omega|_{r=r_{\rm
th},\theta=\pi/2}, $ being the equatorial angular velocity of
rotation of the wormhole throat. Without loss of generality we may
suppose that $\Omega_{\rm th}>0$, i.e. the throat is rotating in
the positive direction. Assume now that the following condition is
fulfilled:
\begin{equation}\label{medl}
R_{\rm th}\Omega_{\rm th}\ll c,
\end{equation}
where $c$ is the velocity of light.  The condition \Ref{medl}
means that the linear velocity of rotation of the throat is much
less than $c$. Further, we will consider an {\em approximation of
slow rotation} with the small dimensionless parameter $\lambda=
R_{\rm th}\Omega_{\rm th}/c$. In this approximation components of
the metric \Ref{metric}, describing the rotating wormhole, should
just slightly differ from respective components of the static
metric \Ref{stat.m}. Following the procedure given in
\cite{Hartle}, we represent the metric functions $A$, $B$, $R$,
$\Omega$ and the field $\Phi$ as an expansion in terms of powers
of $\lambda$:
\begin{eqnarray}\label{razl-w}
\Omega&=&\lambda \omega+O(\lambda^{3}),\\
A&=&A_0(1+\lambda^{2}\alpha)+O(\lambda^{4}),\label{razl-a}\\
B&=&B_0(1+\lambda^{2}\beta)+O(\lambda^{4}),\\
R&=&R_0(1+\lambda^{2}\rho) +O(\lambda^{4}),\\
\Phi&=&\Phi_0(1+\lambda^{2}\phi)+O(\lambda^{4}), \label{razl-phi}
\end{eqnarray}
where $A_0$, $B_0$, $R_0$, and $\Phi_0$ are zero order solutions
corresponding to the unperturbed static spherically symmetric
configuration (\ref{stat.m}), \Ref{stat.phi}:
\begin{equation}\label{zeroordersol}
A_0=e^{2u},\quad B_0=e^{-2u},\quad R_0^2=e^{-2u}(r^2+a^2),\quad
\Phi_0={\frac{(m^2+a^2)^{1/2}}{2\pi^{1/2}\, m}}\,u(r).
\end{equation}
It is necessary to emphasize that the substitution
$\lambda\to-\lambda$ (or, equivalently,
$\Omega_{th}\to-\Omega_{th}$) merely corresponds to the rotation
in the opposite direction. It is obvious that in this case the
angular velocity $\Omega$ is also changing its sign,
$\Omega\to-\Omega$, while the functions $A$, $B$, $R$, and $\Phi$
do not depend on the direction of rotation. Mathematically, this
means that $\Omega$ is an odd function of $\lambda$, i.e.
$\Omega(-\lambda)=-\Omega(\lambda)$, and the others are even, i.e.
$A(-\lambda)=A(\lambda)$, etc. Therefore, the expansion
\Ref{razl-w} for $\Omega$ contains only odd powers of $\lambda$,
while the other expansions (\ref{razl-a}-\ref{razl-phi}) contain
only even powers of $\lambda$.

Substituting the expansions (\ref{razl-w}-\ref{razl-phi}) into the
field equations \Ref{einstein},\Ref{eqmo} and collecting terms
with similar powers of $\lambda$ yields a sequence of $n$-th order
equations with $n$ corresponding to powers of $\lambda$. The
zeroth order field equations describe the static spherically
symmetric configuration (\ref{stat.m}), \Ref{stat.phi}. Cutting
the sequence of equations on a definite $n$ corresponds to the
$n$-th order approximation of the theory.

\subsection{The first order approximation}
Rotating wormholes in the first order approximation have been
studied in \cite{KasSus}. As is seen from Eqs.
(\ref{razl-w})--(\ref{razl-phi}), in this approximation the
functions $A$, $B$, $R$, $\Phi$ remain to be unperturbed,
while the angular velocity
$\Omega$, being initially equal to zero, takes the form
$\Omega=\lambda\omega$. The only nontrivial equation for $\omega$
reads
\begin{equation}\label{equ}
-\frac{1}{\sin^3\theta} \partial_{\theta}
[\sin^3\theta\,\partial_{\theta}\omega]=
(r^2+a^2)\partial_{r}^2\omega+4(r-m)\partial_{r}\omega.
\end{equation}
Its solution, having a natural physical meaning, is
\begin{equation}\label{omegasol}
\omega(r)=\frac{\omega_0(\mu)}{a}
\left[1-e^{4u(r)}\left(1+\frac{4m(r+2m)}{r^2+a^2}\right)\right].
\end{equation}
with
\beq
\omega_0(\mu)=[1-e^{-2\pi \mu}(1+8\mu^2)]^{-1},
\eeq
where $\mu\equiv m/a$ is a dimensionless mass parameter. This
solution describes a slowly rotating wormhole with the angular
velocity $\Omega=\lambda\omega$ and the angular momenta $J_{\pm}$:
\begin{equation}\label{J}
J_\pm=\textstyle\frac43 \mu\omega_0 q_\pm(a^2+4m^2),\quad q_+=1,\
q_-=e^{-4\pi\mu},
\end{equation}
which are defined from asymptotics
$\omega={2J_\pm}{|r|^{-3}}+O(|r|^{-4})$ at $r\to\pm\infty$.

Stress that the first order approximation lets only to determine
the angular velocity $\Omega$ of the wormhole's rotation and, as a
consequence, reveal interesting features in a motion of test
particles and a propagation of light in the rotating wormhole
spacetime \cite{KasSus}.
At the same time, since the metric functions $A$, $B$ and $R$ and
the scalar field $\Phi$ remain to be unperturbed in the first
order approximation one cannot answer a number of important
questions. For example, how does the rotation changes such
characteristics of the static wormhole as its mass and throat's
radius? How does the value of violation of the null energy
condition depend on the wormhole's rotation? To answer these
questions, hereafter we will consider the second order
approximation.

\subsection{The second order approximation}
Substituting the expansions (\ref{razl-w})--(\ref{razl-phi}) into
the Einstein equations (\ref{einstein}) and the scalar field
equation \Ref{eqmo} and collecting $\lambda^2$-terms yields the
following system of equations for $\alpha$, $\beta$, $\rho$, and
$\phi$:
\begin{equation}\label{EqSecApr_tt}
(r^2+a^2)\pd_{r}^2\alpha+(\pd_{\theta}^2\alpha
+\cot\theta\,\pd_{\theta}\alpha)+(2r+m)\pd_r\alpha
-m(\pd_r\beta-4\pd_r\rho)=
e^{-4u}\omega'^2(r^2+a^2)^2\sin^2\theta,
\end{equation}
\begin{eqnarray}
(r^2+a^2)(\pd_{r}^2\alpha+4\pd_{r}^2\rho)+(\pd_{\theta}^2\beta+\cot\theta\,\pd_{\theta}\beta)
+3m\pd_r\alpha+(2r-m)(4\pd_r\rho-\pd_r\beta) \nonumber &&\\
=e^{-4u}\omega'^2(r^2+a^2)^2\sin^2\theta+\displaystyle\frac{8(m^2+a^2)}{m}\pd_r(u\phi),&&\label{EqSecAp_rr}
\end{eqnarray}
\begin{eqnarray}
&&(r^2+a^2)(\pd_{r\theta}^2\alpha+2\pd_{r\theta}^2\rho)-r(\pd_{\theta}\alpha+\pd_{\theta}\beta)+2m\pd_{\theta}\alpha=
\frac{4(m^2+a^2)}{m}u\pd_{\theta}\phi,\label{EqSecAp_rtheta}
\end{eqnarray}
\begin{equation}\label{EqSecAp_thetatheta}
2(r^2+a^2)\pd_r^2\rho+2(\pd_{\theta}^2\rho+\cot\theta\,\pd_{\theta}\rho)
+\pd_{\theta}^2\alpha+\pd_{\theta}^2\beta+
(r-m)(\pd_r\alpha-\pd_r\beta)+4(2r-m)\pd_r\rho-2\beta+4\rho =0,
\end{equation}
\begin{eqnarray}\label{EqSecAp_phiphi}
2(r^2+a^2)\pd_r^2\rho+2(\pd_{\theta}^2\rho+\cot\theta\,\pd_{\theta}\rho)
+\cot\theta\,(\pd_{\theta}\alpha+\pd_{\theta}\beta)
+(r-m)(\pd_r\alpha-\pd_r\beta)+4(2r-m)\pd_r\rho&&\nonumber\\
-2\beta+4\rho
 =-e^{-4u}\omega'^2(r^2+a^2)^2\sin^2\theta,&&
\end{eqnarray}
\begin{equation}\label{EqSecApr_kg}
2u(r^2+a^2)\pd_r^2\phi+2u(\pd_{\theta}^2\phi+\cot\theta\,\pd_{\theta}\phi)+
4(ru+m)\pd_r\phi+m(\pd_r\alpha-\pd_r\beta+4\pd_r\rho)=0,
\end{equation}
where $\pd_r\alpha=\pd\alpha/\pd r$,
$\pd_r^2\alpha=\pd^2\alpha/\pd r^2$, etc. One may simplify the
system (\ref{EqSecAp_rr})--(\ref{EqSecApr_kg}) noting that Eq.
\Ref{EqSecAp_rtheta} can be integrated straightforwardly resulting
in
\begin{equation}\label{EqSecAp_rtheta_int}
(r^2+a^2)(\pd_r\alpha+2\pd_r\rho)-r(\alpha+\beta)+2m\alpha=
\displaystyle\frac{4(m^2+a^2)}{m}u\phi+f_1,
\end{equation}
where $f_1(r)$ is an arbitrary function of $r$. Also, it will be
convenient to consider their combinations
[\Ref{EqSecAp_thetatheta}$+$\Ref{EqSecAp_phiphi}] and
[\Ref{EqSecAp_thetatheta}$-$\Ref{EqSecAp_phiphi}] instead of
equations \Ref{EqSecAp_thetatheta} and \Ref{EqSecAp_phiphi}:
\begin{eqnarray}\label{addeq}
&4(r^2+a^2)\pd_r^2\rho+4(\pd_{\theta}^2\rho+\cot\theta\,\pd_{\theta}\rho)+
(\pd_{\theta}^2\alpha+\cot\theta\,\pd_{\theta}\alpha)
+(\pd_{\theta}^2\beta+\cot\theta\,\pd_{\theta}\beta)
+2(r-m)(\pd_r\alpha-\pd_r\beta)&\\&+8(2r-m)\pd_r\rho-4\beta+8\rho
 =-e^{-4u}\omega'^2(r^2+a^2)^2\sin^2\theta,&\nonumber
\end{eqnarray}
\begin{equation}\label{subtreq}
\pd_{\theta}^2\alpha+\pd_{\theta}^2\beta
-\cot\theta\,(\pd_{\theta}\alpha+\pd_{\theta}\beta)
=e^{-4u}\omega'^2(r^2+a^2)^2\sin^2\theta.
\end{equation}
Then, integrating Eq. \Ref{subtreq} yields
\begin{equation}\label{subtreqint}
\alpha+\beta=\textstyle\frac14
e^{-4u}\omega'^2(r^2+a^2)^2(2\cos^2\theta-1) +f_3\cos\theta+f_2,
\end{equation}
where $f_2(r)$ and $f_3(r)$ are arbitrary functions of $r$. Taking
into account that the rotating wormhole configuration should
possess the symmetry $\theta\to\pi-\theta$ one should set
$f_3(r)\equiv 0$. Finally, the equations \Ref{EqSecApr_tt},
\Ref{EqSecAp_rr}, \Ref{EqSecApr_kg}, \Ref{EqSecAp_rtheta_int},
\Ref{addeq} and \Ref{subtreqint} form the system to be solved.
Note that only four equations of this system are independent since
the Bianchi identity $\nabla_\mu G^{\mu}_{\nu}=0$ and the
conservation law $\nabla_\mu T^\mu_{\nu}=0$ take place.

\begin{figure}[h]
 \centerline{\includegraphics[width=7cm]{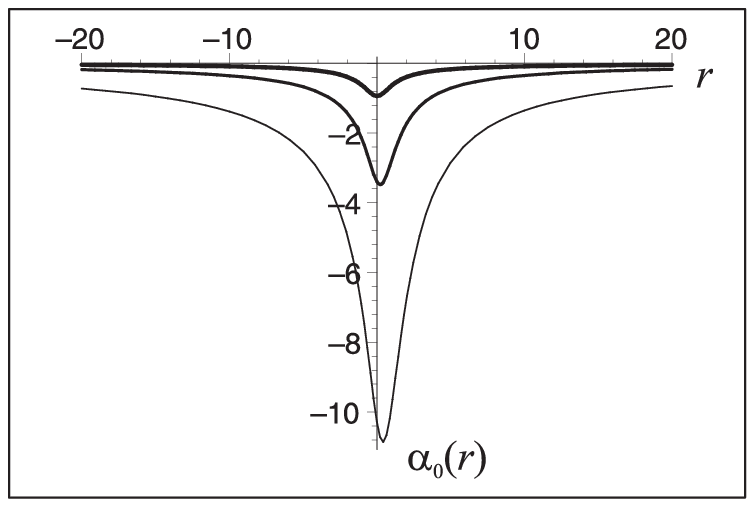}
 \includegraphics[width=7cm]{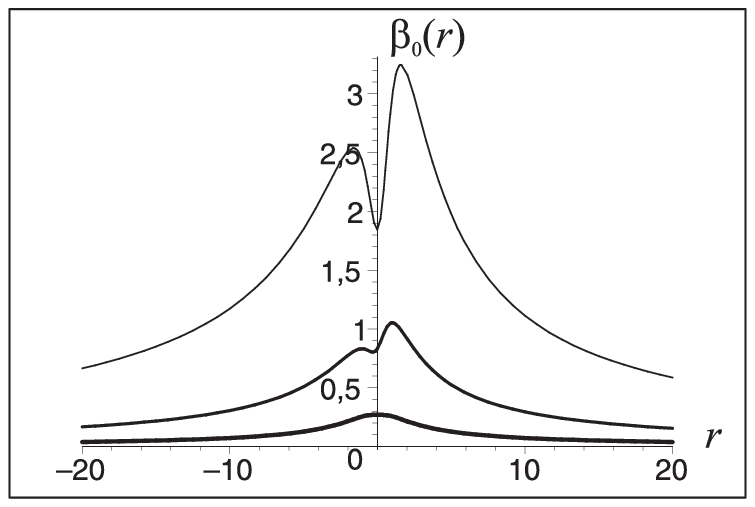}}
 \includegraphics[width=7cm]{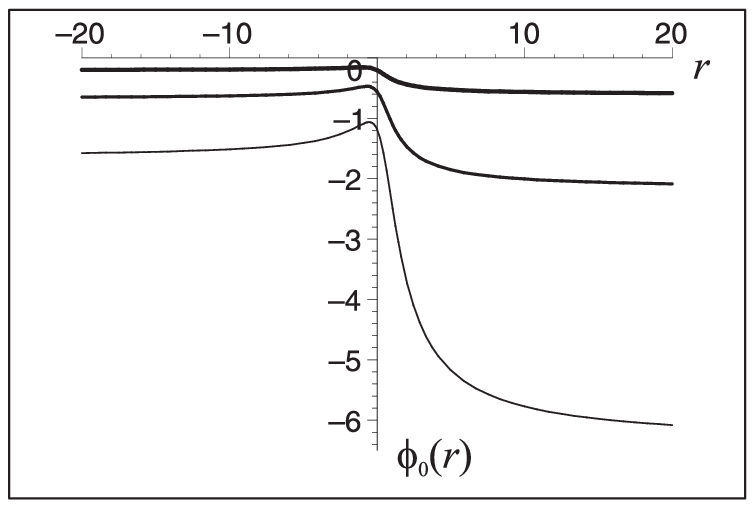} \caption{The graphs of
 $\alpha_0(r)$, $\beta_0(r)$, and $\phi_0(r)$ for $a=1$.
 The thick, middle and thin curves correspond to $m=0$, $0.5$,
 $1$, respectively.
 \label{fig1}}
\end{figure}

\begin{figure}[h]
 \centerline{\includegraphics[width=7cm]{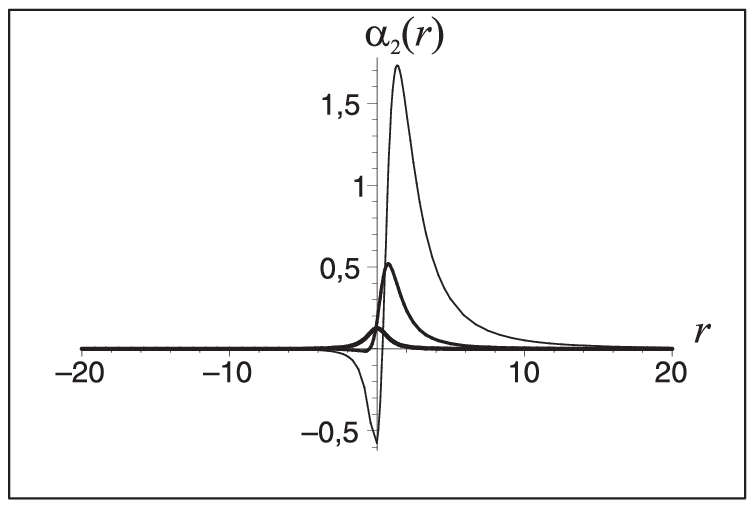}
 \includegraphics[width=7cm]{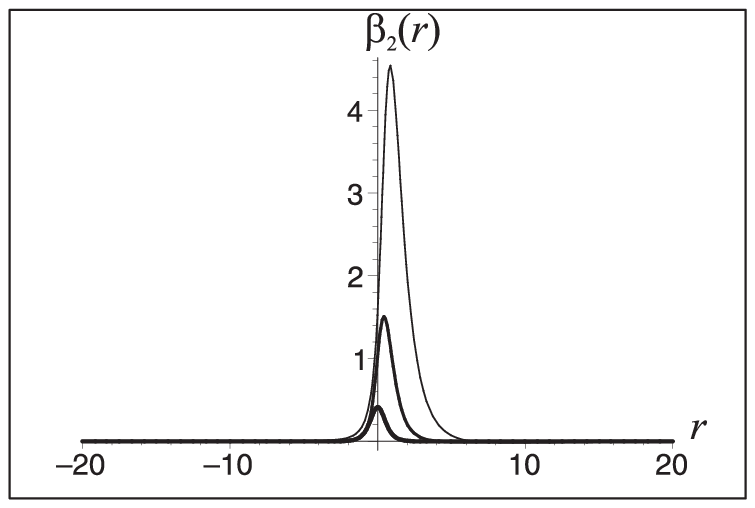}}
 \centerline{\includegraphics[width=7cm]{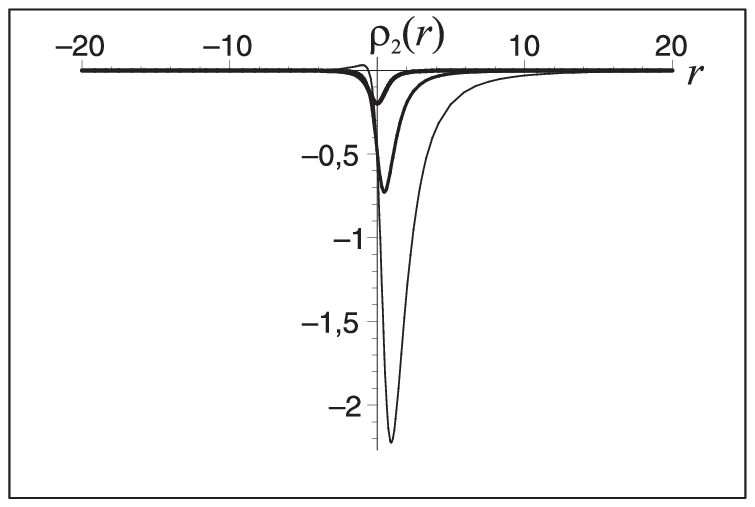}
 \includegraphics[width=7cm]{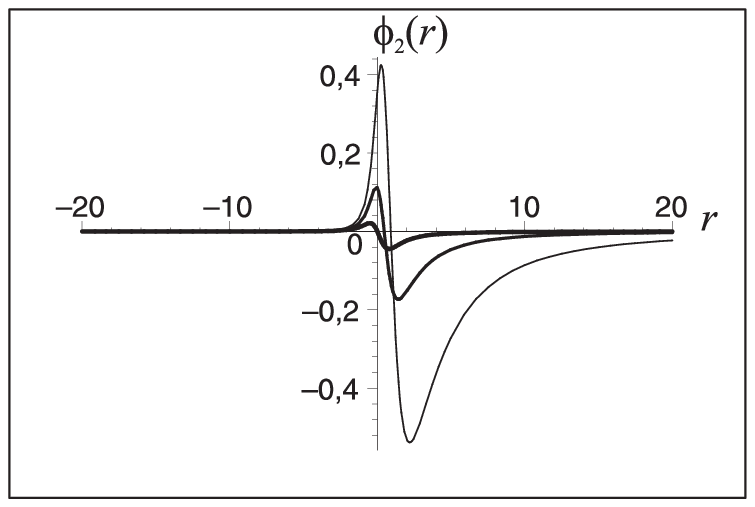}} \caption{The graphs of
 $\alpha_2(r)$, $\beta_2(r)$, $\rho_2(r)$ and $\phi_2(r)$ for $a=1$.
 The thick, middle and thin curves correspond to $m=0$, $0.5$,
 $1$, respectively. \label{fig2}}
\end{figure}

To solve the system of field equations we will follow Ref.
\cite{Hartle} and expand the functions $\alpha$, $\beta$, $\rho$,
and $\phi$ in spherical harmonics:\footnote{Spherical harmonics or
Legendre polynomials $P_{n}(\theta)$ obey the equation (see, for
example, \cite{AbrSte})
$$
\frac{d^2P_n}{d\theta^2}+\cot\theta\,\frac{dP_n}{d\theta}=-n(n+1)P_n.
$$
They satisfy the Rodrigues' formula
$$
P_n(\theta)=\frac{(-1)^n}{2^n
n!}\left(\frac{d}{d\cos\theta}\right)^n\{[1-\cos^2\theta]^n\}.
$$
In particular, $P_{0}(\theta)=1$, $P_1(\theta)=\cos\theta$,
$P_{2}(\theta)={\textstyle\frac{3}{2}\cos^2\theta-\frac12}$. Note
that $P_n(\theta)=(-1)^n P_n(\pi-\theta)$, i.e. even spherical
harmonics (with even numbers $n$) are symmetric with respect to
the equatorial plane $\theta=\frac{\pi}{2}$, while odd ones are
antisymmetric.}
\bs\label{expansionP}
\bea
&&\alpha(r,\theta)=\alpha_{0}(r)+\alpha_{2}(r)P_2(\theta)+...,\\
&&\beta(r,\theta)=\beta_{0}(r)+\beta_{2}(r)P_2(\theta)+...,\\
&&\rho(r,\theta)=\rho_{0}(r)+\rho_{2}(r)P_2(\theta)+...,\\
&&\phi(r,\theta)=\phi_{0}(r)+\phi_{2}(r)P_2(\theta)+....
\eea \es
We would like to stress that the expansions \Ref{expansionP}
contain only {\em even} spherical harmonics being symmetric with
respect to the equatorial plane $\theta=\frac{\pi}{2}$. Another
convenient simplification of the metric may be made here.
Transformations of the type $r\to f(r)$ do not change the form of
the metric \Ref{metric}. Such a coordinate transformation may,
therefore, be used to provide the additional condition
\beq
\rho_{0}(r)=0.
\eeq
This will be assumed in the following. Now substituting Eqs.
\Ref{expansionP} into (\ref{EqSecApr_tt}), (\ref{EqSecAp_rr}),
(\ref{EqSecApr_kg}), (\ref{EqSecAp_rtheta_int}), (\ref{addeq}) and
(\ref{subtreqint}) we find
\bea\label{E1}
(r^2+a^2)\alpha_{n}''+(2r+m)\alpha_{n}'-m(\beta_{n}'-4\rho_{n}')-n(n+1)\alpha_{n}
&=&{\textstyle\frac23}
e^{-4u}(r^2+a^2)^2\omega'^2(\delta_{n0}-\delta_{n2}), \\
\label{E2}%
(r^2+a^2)(\alpha_{n}''+4\rho_{n}'')+3m\alpha_{n}'
-(2r-m)(\beta_{n}'-4\rho_{n}')-n(n+1)\beta_{n}&=&{\textstyle\frac23}
e^{-4u}(r^2+a^2)^2\omega'^2(\delta_{n0}-\delta_{n2})\nonumber\\
&&+8\frac{m^2+a^2}{m}(u\phi_{n})',\\
\label{E3}
(r^2+a^2)(\alpha_{n}'+2\rho_{n}')-(r-2m)\alpha_{n}-r\beta_{n}&=&
4\frac{m^2+a^2}{m}u\phi_{n}+f_1\delta_{n0},\\
\label{E4} 4(r^2+a^2)\rho_{n}''
+8(2r-m)\rho_{n}'+2(r-m)(\alpha_{n}'-\beta_{n}')+8\rho_{n}
-4\beta_{n}\phantom{-4\beta_{n}}&&\nonumber\\
-n(n+1)(4\rho_{n}+\alpha_{n}+\beta_{n})&=& -\frac23
e^{-4u}(r^2+a^2)^2\omega'^2(\delta_{n0}-\delta_{n2}),\\
\label{E5} \alpha_{n}+\beta_{n}&=&\textstyle\frac13
e^{-4u}(r^2+a^2)^2\omega'^2\delta_{n2}+f_2\delta_{n0},\\
\label{kg}
2u(r^2+a^2)\phi_{n}''+4(ur+m)\phi_{n}'-2n(n+1)u\phi_{n}+m(\alpha_{n}-\beta_{n}+4\rho_{n})'&=&0,
\eea
where a prime means the derivative with respect to
$r$, $n=0,2,...$, and $\delta_{n\tilde n}$ is the Kronecker delta.
Note that only the $n=0$ and $n=2$ equations involve the angular
velocity $\omega$. The coefficients in the expansion of $\alpha$,
$\beta$, $\rho$, and $\phi$ with $n\ge4$ must, therefore, vanish
since they vanish when the wormhole is not rotating, i.e.
\beq
\alpha_n=\beta_n=\rho_n=\phi_n\equiv0,\quad n\ge4.
\eeq
This reduction in the number of values of $n$ from infinity to 2
is the central simplification of the slow rotation approximation.
In place of a system of partial differential equations one now
only has ordinary differential equations for the seven unknown
functions $\alpha_0$, $\beta_0$, $\phi_0$, and $\alpha_2$,
$\beta_2$, $\rho_2$, $\phi_2$.
A solution of these equations in an explicit analytic form is
given in the appendix. Generally speaking, this solution depends
on several constants of integrations. However, their values are
fixed if one assumes that the perturbations are everywhere regular
and obey the natural boundary conditions:
\bea
&\alpha_0|_{r\to\pm\infty}=0,\quad
\beta_0|_{r\to\pm\infty}=0,\quad \phi_0|_{r\to\pm\infty}=const,&
\nonumber\\
&\alpha_2|_{r\to\pm\infty}=0,\quad
\beta_2|_{r\to\pm\infty}=0,\quad \rho_2|_{r\to\pm\infty}=0,\quad
\phi_2|_{r\to\pm\infty}=0.&
\eea
These conditions guarantee that there is no rotation far from the
wormhole throat. Note that $\phi_0(r)$ tends to a constant as
$r\to\pm\infty$ because the action \Ref{action} is invariant with
respect to the shift $\Phi\to\Phi+{const}$. In Figs. \ref{fig1}
and \ref{fig2} we give graphical representation for $\alpha_0$,
$\beta_0$, $\phi_0$, and $\alpha_2$, $\beta_2$, $\rho_2$,
$\phi_2$.

\section{Analysis of the solution\label{analysis}}   %
Properties of the non-rotating wormhole spacetime with the metric
\Ref{stat.m} are only determined by two parameters $m$ and $a$,
where $m$ represents the wormhole mass measured by a distant
observer located at $r=\infty$, and $a$ determines the radius of
the wormhole throat. As was shown in the previous section, in
addition to $m$ and $a$ the rotating wormhole solution depends on
the parameter $\lambda=R_{\rm th}\Omega_{\rm th}/c$ being a
dimensionless linear velocity of rotation of the wormhole throat.
In this section we will discuss the problem: How do
characteristics of the rotating wormhole differ from those of the
non-rotating one possessing the same parameters $m$ and $a$?

\subsection{The mass of rotating wormhole}
\begin{figure}[h]
 \centerline{\includegraphics[width=7cm]{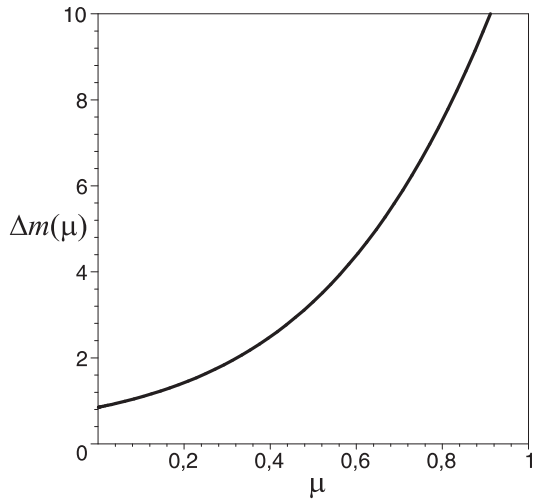}} \caption{The graph of
 $\Delta m(\mu)$ for $a=1$. \label{dmass}}
\end{figure}

To find the rotating wormhole mass $M$, we should consider the
limit $g_{tt}|_{r\to \infty}\to-(1-2M/r)$. In the second order
approximation we have found
$g_{tt}=A(r,\theta)=-e^{2u(r)}\{1+\lambda^2[\alpha_0(r)+\alpha_2(r)
P_2(\theta)]\}$. Note that
$\alpha_2(r)|_{r\to\infty}\sim{r^{-2}}$, hence the $n=2$ solution
does not give any contribution in $M$. Taking into account the
following asymptotical property:
\bea
\alpha_0(r)|_{r\to\infty}&=&-\frac{2\Delta m}{r}+O(r^{-2}),
\eea
with
\beq
\Delta m(\mu)=a\omega_0^2(\mu) \left[
\frac{2(1+10\mu^2)(3\mu\pi+(u(\mu)+1) (
e^{-4\mu\pi}-1))}{3\pi}+\frac{2\mu e^{4u(\mu)}(34\mu^4-\mu^2+1)
}{3(1+\mu^2)}-16\mu^3\right],
\eeq
where $\mu=m/a$ is a dimensionless mass parameter, we obtain
\beq
M=m+\lambda^2\Delta m(\mu),
\eeq
The value of $\lambda^2\Delta m(\mu)$ characterizes the difference
between the rotating and non-rotating wormhole masses $M$ and $m$,
respectively. In Fig. \ref{dmass} the graph of $\Delta m(\mu)$
versus $\mu$ is shown. Note that $\Delta m(\mu)$ is positive for
all $\mu$ and is the greater the greater $\mu$. It is worth to
stress that $\Delta m(\mu)$ is not equal to zero in case $\mu=0$:
\beq\label{dm0}
\Delta m(0)=\frac{8a}{3\pi}.
\eeq
The case $\mu=0$ or, equivalently, $m=0$ corresponds to the
massless non-rotating wormhole. The non-zero value $\Delta m(0)$
means that there do not exist massless rotating wormholes.

\subsection{The NEC violation}
\begin{figure}[h]
 \centerline{\includegraphics[width=7cm]{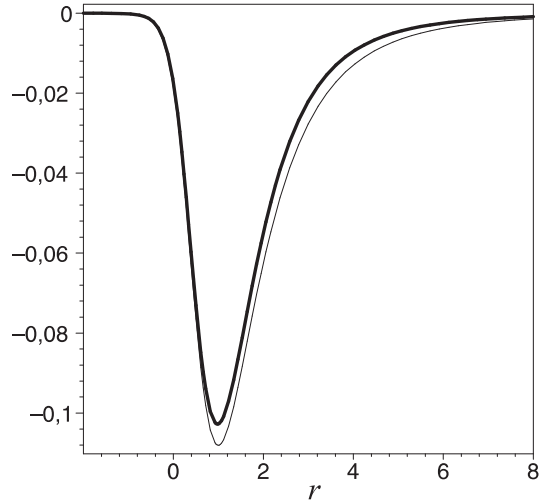}} \caption{The graphs
 of $\Xi(r)$ (thick line) and $\Xi_0(r)$ (thin line) for $a=1$.
 \label{NEC}}
\end{figure}

A violation of the null energy condition (NEC) in the vicinity of
the wormhole throat is an essential feature of wormhole physics
\cite{MorTho,VisserBook,HocVis}.
The NEC reads $T_{\mu\nu}k^\mu k^\nu\ge0$, where $T_{\mu\nu}$ is
the stress-energy tensor, and $k^\mu$ is a null vector. Using the
Einstein equations, it can be represented in the geometrical form
${R}_{\mu\nu}k^\mu k^\nu\ge0$, where ${R}_{\mu\nu}$ is the Ricci
tensor. To analyze the NEC in the rotating wormhole spacetime with
the metric \Ref{metric}, we choose
$k^{\mu}=(A^{-1/2},B^{-1/2},0,\Omega A^{-1/2})$ and introduce the
value $\Xi=R_{\mu\nu}k^\mu k^\nu$. In the second order
approximation $\Xi$ takes the following form
\beq
\Xi(r,\theta)=\Xi_0(r)+\lambda^2[\xi_0(r)+\xi_2(r) P_2(\theta)],
\eeq
where
\beq
\Xi_0(r)=-2e^{2u(r)}\frac{m^2+a^2}{(r^2+a^2)^2},
\eeq
and
\bea
\xi_0(r)&=&\frac{e^{2u(r)}}{(r^2+a^2)^2}\Big[2\beta_0(m^2+a^2)
+(r-m)(r^2+a^2)(\alpha_0'+\beta_0')\Big],\\
\xi_2(r)&=&\frac{e^{2u(r)}}{(r^2+a^2)^2}\Big[2\beta_2(m^2+a^2)
+(r^2+a^2)[-3(\alpha_2-\beta_2)+(r-m)(\alpha_2'+\beta_2'-4\rho_2')
-2\rho_2''(r^2+a^2)]\Big].
\eea
The value $\Xi_0$ characterizes the configuration without
rotation. As is seen, $\Xi_0$ is everywhere negative, hence the
NEC is violated in the whole spacetime of the non-rotating
wormhole. As for rotating wormholes, it will be convenient to
average the quantity $\Xi(r,\theta)$ over all directions:
\bea
\Xi(r)&=&\frac{1}{4\pi}\int_0^{2\pi}\!\!\!\int_0^{\pi}\Xi(r,\theta)\sin\theta d\theta d\varphi\nonumber\\
&=&\Xi_0(r)+\frac{\lambda^2
e^{2u(r)}}{(r^2+a^2)^2}\Big[2\beta_0(m^2+a^2)
+(r-m)(r^2+a^2)(\alpha_0'+\beta_0')\Big].
\eea
In Fig. \Ref{NEC} the quantities $\Xi(r)$ and $\Xi_0(r)$ are shown
together. It is seen that the value of $\Xi(r)$ is everywhere
negative but greater than $\Xi_0(r)$. This means that the NEC
violation in the rotating wormhole spacetime is weaker than that
in the non-rotating one.

\section{Summary\label{conc}}   %
We have constructed a solution describing slow rotating wormholes
in general relativity with the scalar field possessing negative
kinetic energy. The role of a small dimensionless parameter
$\lambda$ plays the ratio of the linear velocity of rotation of
the wormhole's throat and the velocity of light, $\lambda=R_{\rm
th}\Omega_{\rm th}/c$. The field equations have been solved in the
second order approximation with respect to $\lambda$. It is worth
noting that we succeeded in finding a solution in an explicit
analytical form. Its analysis has shown that a mass of a rotating
wormhole is greater than that of a non-rotating one. As a
consequence, this means that rotating wormholes, in contrast to
the non-rotating ones, cannot possess a zero mass. The respective
analysis of the NEC violation in a rotating wormhole spacetime
reveals the fact that it is slightly weaker than that in a
non-rotating one.

\section*{Acknowledgments}
The work was supported by the Russian Foundation for Basic
Research grants No 08-02-00325, 08-02-91307.

\section*{Appendix}
Here we will solve the system of equations (\ref{E1})--(\ref{kg}).
Note that the equations for different values of $n$ are not
coupled together. For this reason, we will consider the equations
for $n=0$ and $n=2$ separately.

\vskip6pt\noindent{\em The case $n=0$.} In this case the equations
\Ref{E3} and \Ref{E5} turn into identities because $f_1(r)$ and
$f_2(r)$ are arbitrary functions. As a result, we obtain
\bea\label{n0-1}
(r^2+a^2)\alpha_{0}''+2r\alpha_{0}'+m(\alpha_0-\beta_{0})'
&=&\textstyle\frac23 e^{-4u}(r^2+a^2)^2\omega'^2,\\
%
\label{n0-2}
(r^2+a^2)\alpha_{0}''+3m\alpha_{0}'-(2r-m)\beta_{0}'&=&
\textstyle\frac23 e^{-4u}(r^2+a^2)^2\omega'^2+ \displaystyle
8\frac{m^2+a^2}{m}(u\phi_{0})',\\
\label{n0-3}
(r-m)(\alpha_{0}-\beta_{0})'-2\beta_{0}&=&-\textstyle\frac13
e^{-4u}(r^2+a^2)^2\omega'^2,\\
\label{n0-4}
2u(r^2+a^2)\phi_{0}''+4(ur+m)\phi_{0}'+m(\alpha_{0}-\beta_{0})'&=&0.
\eea
Note that the first order equation \Ref{n0-3} plays the role of a
differential constraint. A general solution to the system
(\ref{n0-1})--(\ref{n0-4}) can be given explicitly as follows:
\bea
\alpha_0(x)&=&\frac{\omega_0^2}{6(x-\mu)}\bigg\{ C_1+C_2x+C_3xu(x)
\nonumber\\
&&+\frac{8e^{4u(x)}}{x^2+1}\Big[(1+10\mu^2)x^3+24\mu^3x^2+(1+2\mu^2+16\mu^4)x-4\mu^3(1+4\mu^2)
\Big] \bigg\},\\
\beta_0(x)&=&\displaystyle\frac{\omega_0^2}{6\mu(x-\mu)^2}
\bigg\{-(1+x\mu)(C_3\mu u+\mu C_2+C_1)+C_3\mu^2(\mu-x)
\nonumber\\&&+\frac{8\mu
e^{4u(x)}}{x^2+1}\Big[3\mu(1+2\mu^2)x^3-(1+2\mu^2-8\mu^4)x^2+3\mu(1-2\mu^2)x
-48\mu^6-28\mu^4-2\mu^2-1\Big]\bigg\},\\
\phi_0(x)&=&\displaystyle\frac{\omega_0^2}{u(x)}\bigg\{\mu
C_4+u(x)C_5+\frac{1}{12(x-\mu)}\Big[C_1[1+(x-\mu)\arctan{x}]+C_2\mu+C_3\mu
u(x) \Big] \nonumber\\&&\displaystyle+\frac{2\mu
e^{4u(x)}}{3(x-\mu)(x^2+1)}\Big[ 1+16\mu^4-2\mu
(1-8\mu^2)x+(1+4\mu^2)x^2-2\mu x^3 \Big] \bigg\}.
\end{eqnarray}
where $x\equiv r/a$ is a dimensionless radial coordinate,
$\mu=m/a$ is a dimensionless wormhole parameter,
$u(x)=\mu(\arctan{x}-\pi/2)$, and 
$C_k$ ($k=1,...,5$) are constants of integration. Stress that
values of $C_k$ are not free. They are connected by means of the
constraint \Ref{n0-3}. Also, the constants $C_k$ should be chosen
so that to provide an appropriate asymptotical behavior of the
solutions:
\bea
&\alpha_0|_{r\to\pm\infty}=0,\quad
\beta_0|_{r\to\pm\infty}=0,&\\
&\phi_0|_{r\to\pm\infty}=const.&
\eea
Moreover, a choice of $C_k$ should guarantee regularity of the
solutions $\alpha_0(x)$, $\beta_0(x)$, $\phi_0(x)$ at the point
$x=\mu$. Altogether, these conditions let us fix {\em all} values
of the constants $C_k$ as follows
\bea &\displaystyle
C_1=8\mu(1+10\mu^2)\left[1-\frac{u(\mu)}{\pi\mu}\left(e^{-4\pi
\mu}-1\right)\right]-\frac{8\mu
e^{4u(\mu)}(34\mu^4-\mu^2+1)}{\mu^2+1},&\nonumber
\\ &C_2=\displaystyle-8(1+10\mu^2),\quad
C_3=\displaystyle\frac{8}{\pi\mu}(1+10\mu^2)(e^{-4\pi\mu}-1),
&\nonumber\\
&\displaystyle C_4=\frac{32\mu^2-\pi C_1}{24\mu},\quad
C_5=\frac{C_3\mu^3-C_1(1+\mu^2)}{12\mu(1+\mu^2)}.&
\eea

\vskip6pt\noindent{\em The case $n=2$.} The system
(\ref{E1})--(\ref{kg}) now takes the following form:
\bea\label{E1_2}
(r^2+a^2)\alpha_{2}''+(2r+m)\alpha_{2}'-m(\beta_{2}'-4\rho_{2}')-6\alpha_{2}
&=&-2v, \\
\label{E2_2}%
(r^2+a^2)(\alpha_{2}''+4\rho_{2}'')+3m\alpha_{2}'
-(2r-m)(\beta_{2}'-4\rho_{2}')-6\beta_{2}&=&-2v+8\frac{m^2+a^2}{m}(u\phi_{2})',\\
\label{E3_2}
(r^2+a^2)(\alpha_{2}'+2\rho_{2}')-(r-2m)\alpha_{2}-r\beta_{2}&=&
4\frac{m^2+a^2}{m}u\phi_{2},\\
\label{E4_2} 4(r^2+a^2)\rho_{2}''
+8(2r-m)\rho_{2}'+2(r-m)(\alpha_{2}'-\beta_{2}')-16\rho_{2}
-6\alpha_{2}-10\beta_{2}&=& 2v,\\
\label{E5_2} \alpha_{2}+\beta_{2}&=&v,\\
\label{kg_2}
2u(r^2+a^2)\phi_{2}''+4(ur+m)\phi_{2}'-12u\phi_{2}+m(\alpha_{2}-\beta_{2}+4\rho_{2})'&=&0,
\eea
where $v\equiv\frac13 e^{-4u}(r^2+a^2)^2\omega'^2$. A general
solution of the system (\ref{E1_2})--(\ref{kg_2}) is
\begin{eqnarray}
\alpha_2(x)&=&\displaystyle-\omega_0^2\bigg\{\frac{4e^{4u(x)}}{3(x^2+1)^2}\Big[3x^6+12\mu
x^5+x^4(7+22\mu^2)+ 24\mu x^3(\mu^2+1)+x^2(5+36\mu^2+16\mu^4)
\nonumber\\&&\displaystyle +4\mu
x(3+4\mu^2-8\mu^4)-64\mu^6+14\mu^2+1\Big]+\arctan{x}\Big[{\textstyle\frac12}\mu
D_2(9x^2+1)-D_4(3x^2+1)\Big]- \nonumber\\&&\displaystyle +\mu
x^2D_1+\frac{\mu
xD_2(8x^2+7)}{2(x^2+1)}-D_3(3x^2+1)-3xD_4\bigg\},\\
\beta_2(x)&=&\displaystyle\omega_0^2\bigg\{\frac{4e^{4u(x)}}{3(x^2+1)^2}\Big[3x^6+12\mu
x^5+x^4(7+22\mu ^2)+ 24\mu x^3(\mu ^2+1)+x^2(5+36\mu ^2+16\mu ^4)
\nonumber\\&&\displaystyle +4\mu x(3+4\mu ^2-8\mu ^4)-64\mu
^6+14\mu ^2+1+16\mu ^2(1+4\mu ^2)^2\Big]
\textstyle+\arctan{x}\Big[\frac12\mu D_2(9x^2+1)-D_4(3x^2+1)\Big]
\nonumber\\&&\displaystyle +\mu
x^2D_1-3D_4x-3D_3x^2-D_3+{\textstyle\frac72}\mu xD_2+\frac{\mu
x^3D_2}{x^2+1}\bigg\},\\
\rho_2(x)&=&\displaystyle\omega_0^2\bigg\{\frac{2e^{4u(x)}}{3(x^2+1)^2}\Big[3x^6+18\mu
x^5+x^4(7+46\mu ^2) +36\mu x^3(2\mu ^2+1)+x^2(5+76\mu ^2+80\mu ^4)
\nonumber\\&&\displaystyle +2\mu x(9+40\mu ^2+16\mu ^4)-64\mu
^6+32\mu ^4+22\mu ^2+1\Big]\nonumber\\&&\displaystyle
+{\textstyle\frac14}\arctan{x}\Big[D_2(9\mu x^2+6x+\mu
)-2D_4(3x^2+1)\Big] \nonumber\\&&\displaystyle
+{\textstyle\frac12}xD_1(\mu x+1)+{\textstyle\frac14}D_2\left[9\mu
x+4-\frac{2x(x-\mu )}{x^2+1}\right]
-{\textstyle\frac12}D_3(3x^2+1)-{\textstyle\frac32}xD_4\bigg\},\\
\phi_2(x)&=&\displaystyle\omega_0^2\bigg\{\frac{4\mu
^2e^{4u(x)}}{3(x^2+1)^2u(x)}\Big[x^4+4\mu x^3+2x^2(1+4\mu ^2)+8\mu
x(2\mu ^2+1) +32\mu ^4+16\mu ^2+1\Big]\nonumber\\&& -\frac{\mu
\arctan{x}}{4(\mu ^2+1)u(x)}\Big[3x^2(3\mu ^2D_2-D_2-2\mu D_4)+\mu
^2D_2-3D_2-2\mu D_4\Big]- \nonumber\\&&\displaystyle -\frac{\mu
}{4(\mu ^2+1)u(x)} \Big[x^2(2\mu ^2D_1-D_1-6\mu D_3)+x(-3D_2+9\mu
^2D_2-6\mu D_4)-2\mu D_3
\nonumber\\&&\displaystyle-D_1-\frac{2xD_2(\mu
^2+1)}{x^2+1}\Big]\bigg\},
\end{eqnarray}
where $D_k$, ($k=1,...,4$) are constants of integration. Taking
into account the boundary conditions
\bea
&\alpha_2|_{r\to\pm\infty}=0,\quad
\beta_2|_{r\to\pm\infty}=0,\quad \rho_2|_{r\to\pm\infty}=0,
\quad\phi_2|_{r\to\pm\infty}=0,&
\eea
and using the two differential constraints following from the
system (\ref{E1_2})--(\ref{kg_2}) we can fix the values of $D_k$
as follows
\begin{eqnarray}
&D_1=\displaystyle-4\mu (e^{-4\pi \mu }+1),\quad
D_2=\displaystyle\frac{8\mu }{\pi}(e^{-4\pi \mu }-1),&\nonumber\\&
D_3=\displaystyle\frac{2}{3}(1-2\mu ^2)(1+e^{-4\pi \mu }),\quad
D_4=\displaystyle\frac{4}{3\pi}(1-3\mu ^2)(1+e^{-4\pi \mu }).&
\end{eqnarray}

\end{document}